\documentstyle[preprint,prl,oldlfont,aps,epsf]{revtex}
\begin{document}
\title{Reply to the comment by C.Camacho}
\maketitle

In
the recent Letter \cite{LENGTH} we reported the results of simulations
concerning chain length dependence of protein folding time. 
We argued
that first order phase
transition scenario where nucleus size does not depend on chain 
length and  the role of surface energy is played by loop entropy
(which depends logarithmically on chain length)
may explain the observed power law dependence. In the  comment
\cite{Camacho} 
Camacho questions this interpretation by arguing 
that
such a first order-like scenario yields folding time which
is exponential in $N!$. 

It is well-known that free energy barrier at first order transition
is determined by the nucleus size which depends
on interplay of bulk and surface free 
energy and does not depend on system size but rather on deviation from transition temperature. 
For this reason {\em at the transition temperature} 
in infinite {\em homogeneous system} the relaxation time diverges,
because the critical nucleus size diverges. However, below the
transition point the relaxation time of a first order transition is 
 finite and by no means it is exponential in factorial of 
system size. It was made very clear in
\cite{LENGTH} that folding kinetics for each sequence
was studied at the conditions {\em of its  fastest folding}, rather
than at temperature of folding transition. In all cases
the temperature of fastest folding for designed sequences 
was 
markedly below the thermodynamic transition temperature.
Therefore the 
allusion in \cite{Camacho} to the scaling of 
relaxation time at the
point of  transition is not applicable to simulations
reported in \cite{LENGTH}. Moreover, conclusion about 
divergence of relaxation
time at the transition point  is directly 
applicable only to homogeneous nucleation: in heteropolymeric
system, even at the point where free 
  energy of the folded state is equal to that of unfolded, some fragments
of structure are more stable in their folded state than other 
and they
can still serve as finite size nuclei for transition
even at temperature of 
 thermodynamic folding transition. For this reason
simulations do not reveal any singular behavior of folding time
in a wide range of temperatures and stabilities.

The analysis presented in \cite{Camacho_PRL} predicts second-order
folding transition, contradicting the results of numerous  
simulations and analytical theories
\cite{GSW} which all converge that  
folding transition in 3 dimensional proteinlike heteropolymer 
is
a first order one. First, we note   that the second order folding transition
is in dramatic disagreement with  experiment 
which shows that folding transition in the majority of 
relatively small 
(up to about 200 aminoacids) proteins
is first-order like (for finite systems), both in thermodynamics
and in kinetics \cite{Creighton}. 

Second, we point out that
the conclusion about ''second order'' transition is a consequence
of a crucial  
uncontrolled assumption made in \cite{Camacho_PRL}.
The loop closure entropy was assumed in \cite{Camacho_PRL} 
to depend linearly on chain length. This is equivalent to neglecting 
chain connectivity in a polymer model. In order to see 
this, consider two monomers, say number $i$ and $j$ in sequence
which are in contact with each other. This closes the  loop 
of  $j-i-1$ monomers between them. 
If conformational statistics of this loop is Gaussian, 
the entropic cost to bring a monomer $k$, which belongs to the
loop (i.e. $i<k<j$), in contact with monomers $i$ and $j$ is 
$d/2 (\log(k-i)+\log(j-k)-\log(j-i))$, (where d is space dimension), 
i.e it depends on position $k$
of the monomer in sequence. 
This feature of the polymer model gives rise to sequence
specificity, which is a key requirement for a model to be protein-like. 
However, in the model presented in \cite{Camacho_PRL} the 
loop entropy
cost of bringing monomer $k$ in contact with $i$ and $j$
from the previous example is
 $\lambda ((k-i)+(j-k)-(j-i))=0$,
which means that sequence dependence 
does not at all exist in this model. 
This makes the model of Camacho equivalent
to a system of disconnected monomers occupying volume $V \sim N$.
 For this reason the ``transition temperature'' in \cite{Camacho_PRL},
goes to zero in thermodynamic limit, i.e. strictly speaking, 
there is no folding transition 
in this model, at all. 

It was argued in \cite{Camacho_PRL} that neglect of chain connectivity
is a mean-field (MF) approximation equivalent to the one made in theory
of random heteropolymers (RHP) \cite{GO,SG89}. This assertion is incorrect.
In fact chain connectivity 
plays a crucial role in MF theory
of RHP discussed in \cite{SG89}. 
The easiest way to see this is to note that the physics
of RHP, as predicted by MF theory 
dramatically depends on space dimension (which enters the theory
via loop entropy factors) with $d>2$ and $d<2$ cases belonging to different
universality classes (ultrametric landscape for $d<2$ and thermodynamic equivalence to the Random Energy Model at $d>2$ \cite{SG89}). It is also clear that the model of Camacho is equivalent to zero-dimensional 
RHP, which explains its inconsistency 
with basic thermodynamic properties
of 3-dimensional heteropolymers and proteins. 

The earlier theories of RHP are mean field  ones not because
they are based on some arbitrary assumptions but because they neglect fluctuations
of certain order parameters, such as 
replica overlap $Q_{\alpha \beta}$ or microphase separation $m(r)$.
In fact the validity of the MF approximation (Ginsburg number) 
in RHP theory
was analyzed, 
and fluctuational corrections (mainly in one-loop approximation) were
calculated in \cite{SGS}

In summary we note that quantitative coincidence between simulations
and inconsistent theory may be only fortuitous.

\vspace{10pt}
\noindent
A. Gutin V.Abkevich and E. Shakhnovich \\
Harvard University \\ 
Department of Chemistry and Chemical Biology\\
 12 Oxford Street, Cambridge MA 02138

\end{document}